\documentclass{osa-article}

\journal{oe}

\articletype{Research Article}
\usepackage{bm}
\begin{document}
\title{A single-pass Cr:ZnSe amplifier for broadband infrared undulator radiation}

\author{M.B.~Andorf\authormark{1*}, V.A.~Lebedev\authormark{2}, P.~Piot\authormark{2,3,4} }

\address{\authormark{1} Cornell University, Ithaca, NY 14853, USA\\
\authormark{2} Fermi National Accelerator Laboratory, Batavia, IL 60510, USA \\
\authormark{3} Northern Illinois University, Department of Physics, DeKalb, IL 60115, USA \\
\authormark{4} Advanced Photon Source, Argonne National Laboratory, Lemont, IL 60439, USA }

\email{\authormark{*}mba83@cornell.edu} 



\begin{abstract}
An amplifier based on a highly-doped Chromium Zinc-Selenide (Cr:ZnSe) crystal is proposed to increase the pulse energy emitted by an electron bunch after it passes through an undulator magnet. The primary motivation is a possible use of the amplified undulator radiation emitted by a beam circulating in a particle accelerator storage ring to increase the particle beam's phase-space density---a technique dubbed Optical Stochastic Cooling (OSC). This paper uses a simple four energy level model to estimate the single-pass gain of Cr:ZnSe and presents numerical calculations combined with wave-optics simulations of undulator radiation to estimate the expected properties of the amplified undulator wave-packet.
\end{abstract}

\section{Introduction}
Optical Stochastic Cooling (OSC) is a technique proposed to increase the phase-space density of a charged-particle beam circulating in a storage-ring accelerator~\cite{OSC_Mikhailichenko,OSC_Zolotorev}. Specifically, OSC aims to take advantage of the large bandwidth supported by optical amplifiers in the optical and mid-IR regime in order to attain faster cooling times than what can be achieved with conventional stochastic cooling. Thus, OSC is an extension of stochastic cooling originally developed in the microwave band~\cite{Meer,Mohl2,Bramham}. In ordinary stochastic cooling, the granular particle beam information (e.g. average particle position) within a longitudinal (temporal) sample slice of the beam is measured via an electrical ``pickup" which the beam passes through. The resulting electrical signal is the summation of error signals associated to each sample slice. The error signals are processed and amplified and then used to apply corrective kicks on the sampling slices at a later time as the beam passes through the downstream ``kicker" element. This sample-and-kick technique is repeatedly applied turn-by-turn as the beam mixes and circulates in the storage ring resulting in a stochastic cooling of the beam. Note that, owing to the longitudinal dynamics in the ring (synchrotron motion) the sampling slices represent different statistical realizations of the bunch population at each turn. The performance of this cooling technique is ultimately limited by the temporal duration of the sampling slices which is determined by the overall bandwidth of the hardware associated with the pickup and kicker electrodes and amplifier system. State-of-the-art stochastic cooling systems are limited to bandwidths of $\sim 8$~GHz~\cite{Pasquinelli}. 

By contrast, broadband lasing media can support bandwidths of $\sim 100$~THz. The increased bandwidth reduces the temporal width of the longitudinal sampling slice dramatically, thereby enabling a finer correction to the beam. In order to transition to optical wavelengths, the conductive plates typically employed for the pickup and kicker in ordinary stochastic cooling, are replaced with magnetic insertion devices called undulators. An undulator consist of a series of alternating polarity dipole magnets that produces a sinusoidal motion of the particle beam and results in it emitting electromagnetic radiation. The spatial-period of the motion, $\lambda_u$ together with the magnetic-field strength and electron energy, ${\cal E}_o $, determine the wavelength emitted by the particle in the forward direction~\cite{Wiedemann}
\begin{equation}
\lambda_o=\frac{\lambda_u}{2\gamma^2} \left(1+\frac{K^2}{2}\right),
\end{equation}
where $\gamma \equiv \frac{{\cal E}_o}{mc^2}$ is the Lorentz factor (here $m$ and $c$ are respectively the electronic mass and the speed of light), and $K$ is the dimensionless undulator parameter. A particle passing through an undulator consisting of $N_u$ periods radiates a short wave-packet of $N_u$ optical cycles yielding a radiation bandwidth $\Delta\lambda/\lambda_0 \sim 1/N_u$ in the forward direction. However, the bandwidth increases considerably for radiation emitted at small-angles from the forward direction (of the order $\sim 2/\gamma$), as will be detailed later.

\begin{figure}[!h]
	\centering
	\includegraphics*[width=0.90\textwidth]{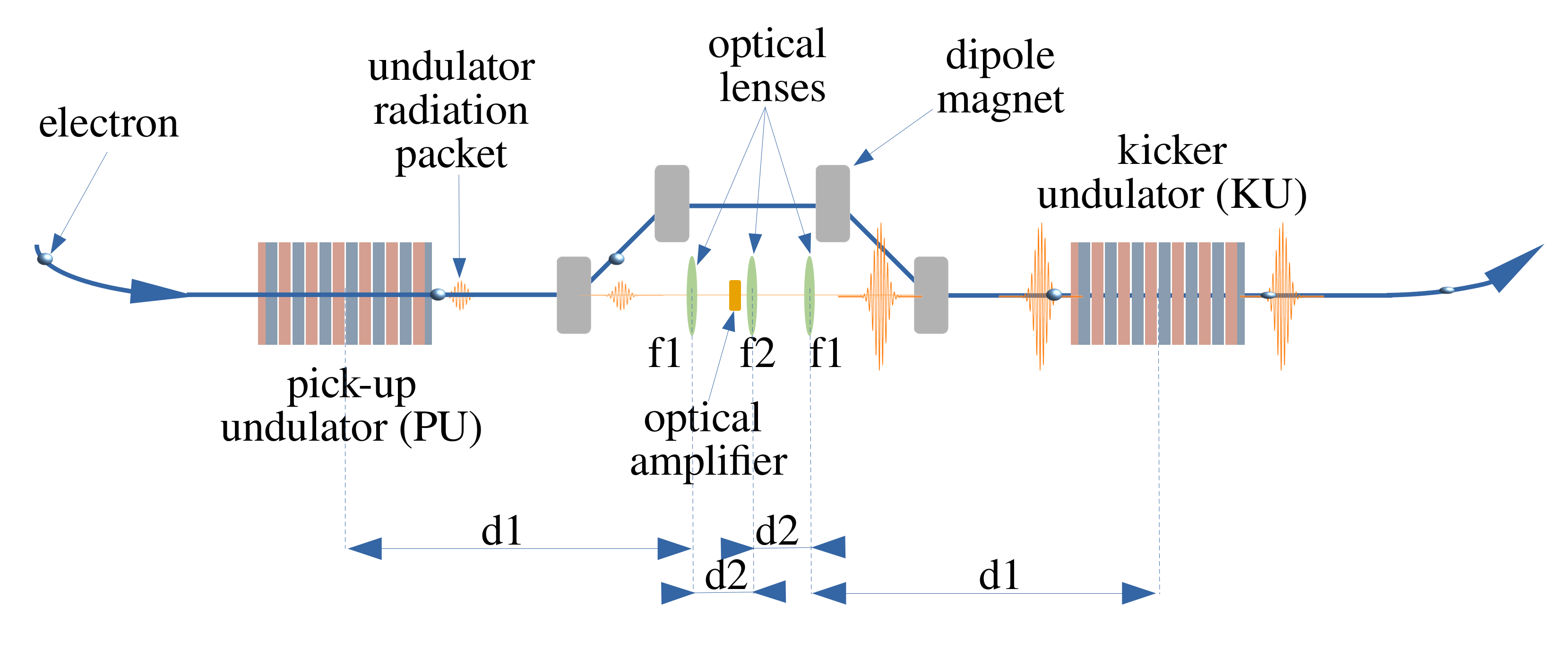}
	\caption{Conceptual overview of the OSC-insersion beamline. The circulating electrons in the storage ring propagate  from the left to right along the blue path; see text for details.}
	\label{fig:oscinsert}
\end{figure}

In OSC, a beam passes through the pickup undulator (PU) to produce an optical pulse. The beam is then separated from the optical pulse via a magnetic bypass chicane, e.g., consisting of four dipole magnets with $(+,-,-,+)$ polarity; see Fig.~\ref{fig:oscinsert}. The chicane temporarily offsets the beam horizontally to allow for the insertion of focusing optics and an amplifier, while also delaying the beam and introducing a linear correlation between the path length associated with a given particle and its initial momentum, relative to a ``reference" particle (an idealized particle which follows precisely the design orbit in the ring). After the pulse is amplified, it is focused into the kicker undulator (KU) where it co-propagates with its parent particle. Each particle will receive an energy kick which is determined by the relative arrival time between optical pulse and particle at the entrance of the KU. In order to maximize the cooling rate, the arrival time of the reference particle is set so that it receives no net kick during the interaction in the kicker. However, a generic particle will be delayed longitudinally by an amount proportional to its momentum deviation, and thus receives a corrective energy kick given by 
\begin{equation}
\Delta \mathcal{E}_p=-\Delta \mathcal{E}\sin(ks_p)
\label{kick1}
\end{equation} 
where $k\equiv 2\pi/\lambda_o$ and $s_p=R_{56}\frac{\Delta p}{p}$ is the longitudinal displacement of the particle from the reference. The coefficient $R_{56}$ is an element of the electron-beam transport matrix from PU to KU centers~\cite{transport}. The process just described results in longitudinal cooling of the particle beam. More generally, by adjusting the accelerator magnetic lattice so that there is horizontal-longitudinal coupling at the PU and KU, simultaneous horizontal and longitudinal damping can be achieved~\cite{OSC_val}.

The energy-kick amplitude, $\Delta \mathcal E$, depends on the PU and KU undulator parameters, the properties (e.g. gain, bandwidth and group velocity dispersion) associated with the optical amplifier and the optical-transport beamline. The kick amplitude ultimately determines how effectively a beam can be cooled with OSC.

An especially important aspect of the cooling system is the damping of particles that have been driven out to large amplitudes, predominantly due to intra-beam~\cite{ibs1,ibs2} and residual-gas~\cite{rgs} scattering processes. Equation~\ref{kick1} indicates that particles with large-amplitudes can experience a delay such that $\psi\equiv ks_p >\pi $. The corresponding energy kick switches signs resulting in particles being displaced toward larger-amplitude orbits (a phenomenon dubbed as ``anti-damping''); see Ref.~\cite{andorf_napac16,zholents_dampingforce}. An analysis of the linear beam optics indicates damping of large-amplitude particles can be accomplished, provided the total delay introduced by the chicane (which is matched to the optical delay set by the amplifier and lenses) is limited to a few millimeters. Unfortunately this seriously constrains the design of the amplifier and prevents the attainment of high-gain amplification. A remedy to increase the number of particles within the cooling-range phase ($\psi < \pi $) is to operate the optical system at longer wavelengths which indicates mid-IR amplifiers may be more suitable for OSC.

\begin{table}[tttt!!!]
    \begin{tabular}{l l l}
        \hline
        \hline
        parameter, symbol & value & unit. \\
        \hline
        undulator parameter, $K$         & 1.04 & -\\
        length,  $L_u$    & 77.4 & cm           \\
        undulator period,  $\lambda_u$    & 11.06 & cm           \\
        number of periods, $N_u$    & 7 & - \\
        on-axis wavelength, $\lambda_o$         & 2.2 & $\mu$m \\
        electron Lorentz factor, $\gamma$    & 195.7 & - \\
        \hline
        \hline
    \end{tabular}
    \caption{Pickup and kicker parameters for the OSC test planned at the IOTA facility with a 100-MeV electron beam.}
    \label{und_params}
\end{table}
The investigation presented in this paper is primarily motivated by a proof-of-principle experiment of the OSC in preparation at Fermilab using a 100-MeV electron beam circulating in a compact ($\sim 40$-m circumference) storage ring called the integrable optics test accelerator (IOTA)~\cite{osc_IOTA}. IOTA was developed to explore new concepts related to the storage of high intensity beams~\cite{Antipov,Shiltsev,Jarvis}. The undulator parameters for the PU and KU are taken to be identical; see Table~\ref{und_params}. While the first stage of the OSC experiment considers a passive-cooling scheme i.e. no amplification~\cite{Andorf_PRAB}, a subsequent phase will demonstrate cooling with an amplifier based a on 1-mm-thick Chromium-doped Zinc-Selenide (Cr:ZnSe) gain medium pumped by a CW Thulium fiber laser with wavelength $\lambda_p=1908$~nm. In addition to being at an advantageous wavelength for OSC, Cr:ZnSe boast a large bandwidth that can support fs-scale pulses\cite{Sorokina2}, making it very suitable for amplifying the broadband radiation generated from an undulator. 

In this paper, we present formulae suitable for computing the amplifier gain and use a wave-optics simulations code to model the amplification over the broad bandwidth of the undulator wave-packet. We find that as a result of the constraints imposed on the amplifier design, the gain is limited to a modest value of ${\cal G}\sim 7$~dB. This gain is sufficient for the  proof-of-principle demonstration planned at the IOTA ring while applications of the OSC technique to cool hadron beams will likely require optical gain of at least 20-dB which may be attainable with a longer Cr:ZnSe crystal.

\section{Gain model for Cr:ZnSe crystal}
\begin{figure}
	\includegraphics[width=0.75\textwidth]{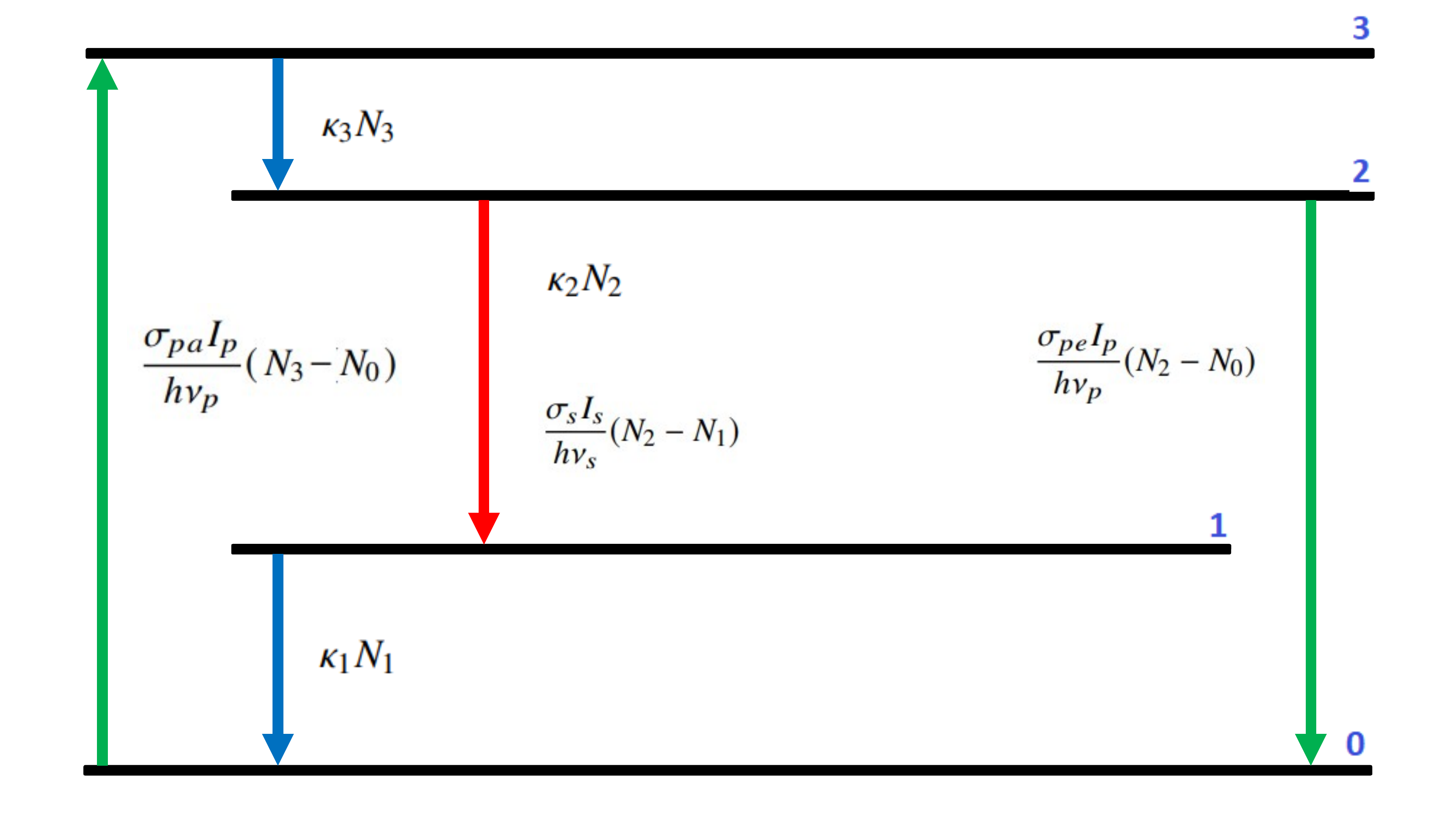}
	\caption[fig:LevelDiagram]{Illustration of the 4-level pumping scheme used to model Cr:ZnSe. The green arrows represents transitions from the pump laser while the red indicates transitions from spontaneous or stimulated emission. Blue arrows indicate radiation-less transitions.}
	\label{fig:EnergyLevel}
\end{figure}

In this section we derive the gain formulae applicable to the regime of undulator-radiation amplification considering a Cr:ZnSe crystal -- modelled as a 4-level gain medium -- subjected to CW pumping  at an intensity $I_p$ with a wavelength of 1908~nm. An illustration is shown in Fig.~\ref{fig:EnergyLevel}. The system of coupled equations governing the evolution of the population density $N_i$ ($i\in[0,3]$) associated with the $i^{th}$ energy state is
\begin{equation}
\begin{split}
&\frac{dN_3}{dt}=-\kappa_3N_3+\frac{\sigma_{pa}I_p}{h\nu_p}(N_0-N_3) \\
&\frac{dN_2}{dt}=\kappa_3N_3-\kappa_2N_2-\frac{\sigma_sI_s}{h\nu_s}(N_2-N_1)-\frac{\sigma_{pe}I_p}{h\nu_p}(N_2-N_0) \\
&\frac{dN_1}{dt}=\kappa_2 N_2-\kappa N_1+\frac{\sigma_sI_s}{h\nu_s}(N_2-N_1) \\
&\frac{dN_0}{dt}=\kappa_1N_1-\frac{\sigma_{pa}I_p}{h\nu_p}(N_0-N_3)+\frac{\sigma_{pe}I_p}{h\nu_p}(N_2-N_0) \\
\label{rates}
\end{split}
\end{equation}
where $\kappa_i$ is the decay rate of the $i^{th}$ energy state, $\nu_{p/s}$ is the photon frequency at the pump/signal wavelength, $\sigma_{pa}$ and $\sigma_{pe}$, respectively, refer to the absorption and emission cross sections,~$\sigma_s$ is the stimulated emission cross-section at the signal wavelength and $I_s$ is its intensity. The emitted radiation from the PU arrives at the entrance of the amplifier in short bunches at the revolution frequency of the storage ring. For the case of IOTA the root-mean-square (RMS) electron bunch duration is on the order of 300~ps and the revolution frequency is 7.5~MHz. For parameters presented in Table~ \ref{und_params} this corresponds to an average power per electron of 60~fW. Given the maximum number of electrons per bunch of $\sim 10^7$, the average input signal power for the amplifier is $\sim 600$~nW and the peak power is only about 300 times greater. We will later show the amplifier requires approximately 40~W of pump power, and so the average power emitted from spontaneous emission is several orders-of-magnitude greater than the input signal. Consequently, the population inversion is not affected by the stimulated emission of the signal pulses during amplification, thereby allowing us to neglect the signal intensity $I_s\simeq 0$ and use a steady-state solution ($\frac{dN_i}{dt}=0$ for all $i \in[0,3]$) to compute the amplifier gain. Note that the rate equations also account for the overlap of emission and absorption cross-sections around 1908~nm.

The pump attenuation is given by
\begin{equation}
\frac{dI_p}{dz}=-I_p\bigg(\sigma_{pa}(N_0-N_3)+\sigma_{pe}(N_0-N_2)\bigg)
\label{initial_di_p_dz}
\end{equation}
and the signal growth by
\begin{equation}
\frac{dI_s}{dz}=I_s\sigma_s(N_2-N_1)
\label{initial_di_s_dz}
\end{equation}
where $z$ is the length inside the crystal and the population densities will depend on $z$.
We assume the decay time from levels 3 and 1 are significantly shorter than 2 so that $N_3\ll N_0$ and $N_1 \ll N_2$. Thus, the first equation in Eq.~\ref{rates} becomes
\begin{equation}
    N_3\approx \frac{I_p\sigma_{pa}}{h\nu \kappa_3}N_0
\end{equation}
which can be used to eliminate $N_3$ from the second equation in Eq.~\ref{rates} Then, using Eq.~\ref{initial_di_p_dz} and Eq.~\ref{initial_di_s_dz} we find an expression that relates the growth of the signal to attenuation of the pump laser along the crystal
\begin{equation}
\frac{1}{I_s}\frac{dI_s}{dz}=-\frac{\sigma_s\tau}{h\nu_p}\frac{dI_p}{dz}
\label{dI_sdz}
\end{equation}
which has the solution
\begin{equation}
{\cal G}=\exp\big(\Delta I_p\frac{\sigma_s\tau}{h\nu_p}\big),
\label{gain}
\end{equation}
where ${\cal G}\equiv I_s/I_{so}$, $\tau$ is the fluorescence decay time of the second excited state, $\Delta I_p\equiv I_P(z=0)-I_p(z=L)>0$ is the absorbed pump intensity after passing through the entire length of the crystal, $L$. Next, to find an expression for the pump attenuation, we note that the majority of the $Cr2+$ ions will occupy either the ground or 2nd excited state. If there are $N_T$ ions in total, then $N_T\approx N_0+N_2$. Using this expression to eliminate $N_2$ from the second equation of Eq.~\ref{rates} yields
\begin{equation}
N_0=N_T\frac{1+I_p\frac{\sigma_{pe}\tau}{h\nu_p}}{1+I_p\frac{\tau}{h\nu_p}(\sigma_{pa}+2\sigma_{pe})}.
\end{equation}
 This expression can be used with Eq.~\ref{initial_di_p_dz} to find the attenuation of the pump laser through the crystal
\begin{equation}.~\label{dipdz_final}
\frac{dI_p}{dz}=-I_pN_T\bigg(\frac{\big(1+I_p\frac{\sigma_{pe}\tau}{h\nu_p}\big)(\sigma_{pa}+2\sigma_{pe})}{1+I_p\frac{\tau}{h\nu_p}(\sigma_{pa}+2\sigma_{pe})}-\sigma_{pe}\bigg),
\end{equation}
which can be numerically integrated and used with Eq.~\ref{gain} to compute the signal gain of the amplifier.
\\
\begin{figure}
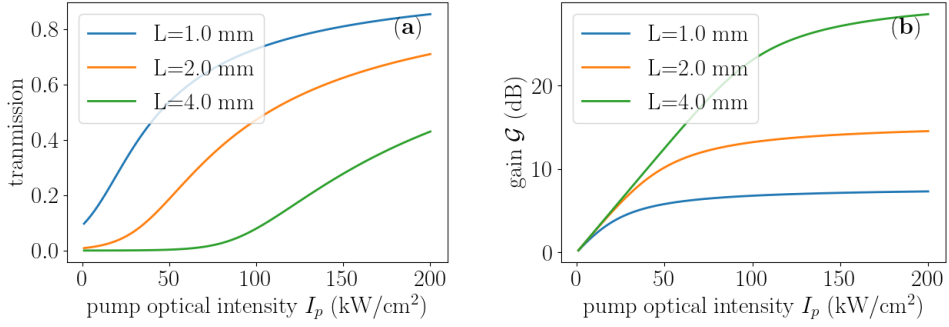

	\centering
	\includegraphics[width=0.49\textwidth]{transmission.pdf}
	\includegraphics[width=0.49\textwidth]{gain.pdf}
	\caption[fig:TGCr:ZnSe]{The transmission (a) and gain (b) as a function of  pump-laser intensity in a  Cr:ZnSe single-pass amplifier   for three different crystal lengths $L$. The transmission an amplification are respectively computed at the  wavelength of $\lambda_p=1908$~nm and $\lambda_o=2.45$~\textmu{m}.}
	\label{fig:gain_and_transmission}
\end{figure}
The gain and transmission for a single-pass through a Cr:ZnSe crystal appears in Fig.~\ref{fig:gain_and_transmission} using values summarized in Table~\ref{xtal_params}. The 1-mm-thick crystal  corresponds to a 1.45-mm optical delay and is the case expected for the active OSC implementation in the IOTA ring. The transmission increases rapidly with the pump-laser  intensity. Such a dependence of the transmission yields a levelling off of the absorbed pump intensity which ultimately limits the attainable gain. At $I_p=125$~kW/cm$^2$ the gain reaches a saturated value of $\sim 7$~dB. For thicker crystals, the onset of the  transmission occurs at higher pump-intensity values resulting in a significant increase in the gain: a 4-mm crystal is expected to provide more than 20~dB of gain. Unfortunately a longer crystal can not be accommodated in the IOTA experiment given the limited optical delay set by the particle delay in the chicane bypass but may have application in another implementation of OSC based on a scalable-delay bypass beamline~\cite{Bergan}. As will be seen below a doping increase which, potentially, could increase the gain is limited by increased thermal stress due to absorbed radiation of pump power. Gain formulas for Cr:ZnSe have also been derived in Ref.~\cite{Berry} and were found to agree with our model to within $\sim 10 \%$
\begin{table}[tttt!!!]
    \begin{tabular}{l l l}
        \hline
        \hline
        parameter, symbol & value & unit. \\
        \hline
        Absorption cross-section at 1908 nm, $\sigma_{pa}$         & 1.0$\times 10^{-18}$ & cm$^2$ \cite{mirov}\\
        Emission cross-section at 1908nm,  $\sigma_{pe}$    & $0.4\times 10^{-18}$ & cm$^2$ \cite{mirov}           \\
        Peak emission cross-section at 2450 nm,  $\sigma_s$    & $1.3\times 10^{-18}$ & cm$^2$ \cite{mirov}        \\
        Fluorescence time, $\tau$    & 5.5  & $\mu$s\cite{mirov} \\
        Total dopant concentration, $N_T$ & 2.0$\times 10^{19}$ & ions/cm$^{3}$\\
        Index of refraction at 2450 nm, $\lambda_s$         & 2.45 & $-$\cite{landolt} \\
        Thermo-optics coefficient, $dn/dT$    & $70\times 10^{-6}$ & $K^{-1}$\cite{Refractive_index_HLI} \\
        Thermal conductivity, $\kappa_t$ & 1.0 & W/cm-K\cite{slack}\\
        Operating temperature & 77 & K\\
        \hline
        \hline
    \end{tabular}
    \caption{Values for Cr:ZnSe parameters used in the presented calculations and simulations.}
    \label{xtal_params}
\end{table}

\section{Transport and amplification of undulator radiation}

The wave-packet emitted by a single electron in the PU has a duration $\tau_{packet} \sim  N_u\lambda_o/c \sim {\cal O}\mbox{(fs)}$ and therefore has a bandwidth which is comparable to the bandwidth of Cr:ZnSe. Consequently, the time domain amplitude growth of the wave packet does not scale with $\sqrt{{\cal G}}$ as it would for a plane-wave tacitly assumed in the above calculations. To make a more realistic estimate of the amplitude growth we use the wave-optics code {\sc Synchrotron Radiation Workshop} (SRW)~\cite{SRW1,SRW2,SRW3} to compute the  emitted radiation field from a single-electron in the PU and propagate it through the optical system, comprising of the amplifier and imaging lenses, to the location of the KU. The simulated field is then used to investigate the interaction between it and a co-propagating electron in the KU.  
\begin{figure}[!h]
	\centering
	\includegraphics*[width=0.49\textwidth]{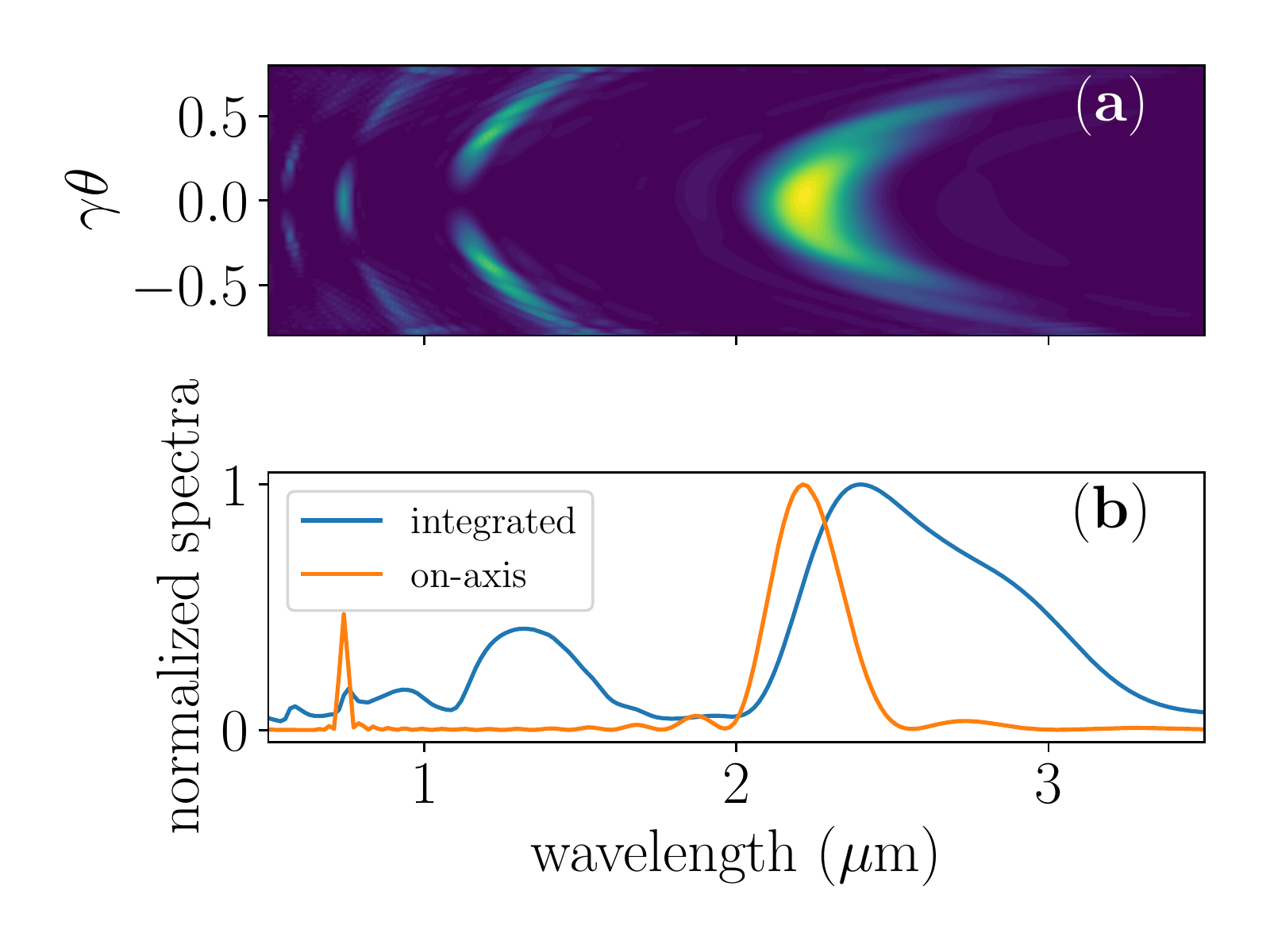}
	\includegraphics*[width=0.49\textwidth]{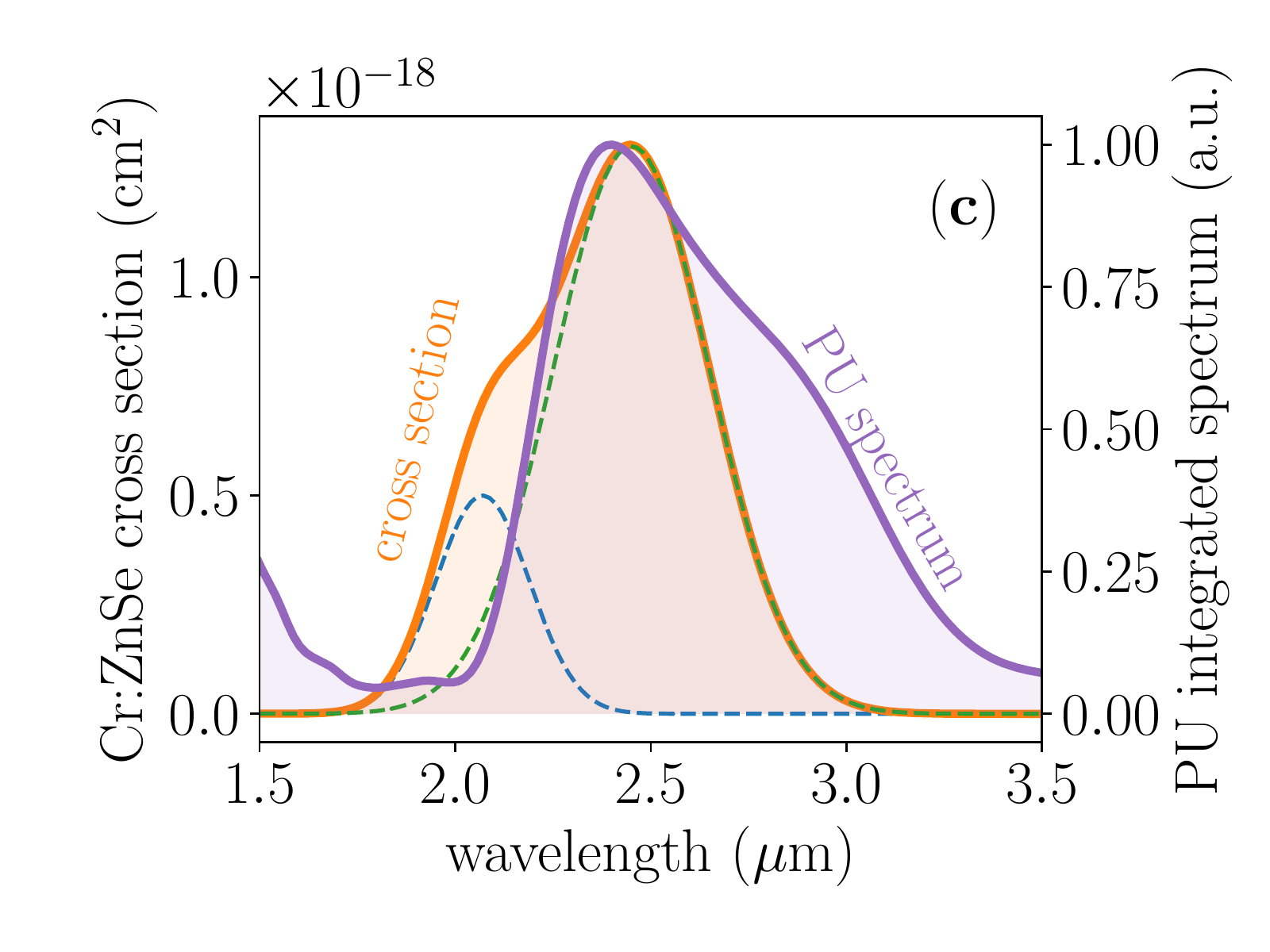}
	\caption{Angle-wavelength radiation fluence (a) with associated on-axis (orange trace) and angle-integrated (blue trace) spectra (b). Comparison of cross-section model for Cr:ZnSe (orange shaded area) with pickup undulator (PU) radiation spectrum (purple shared area) (c) . The dashed lines in (c) represent the two Gaussian distributions used to model the wavelength dependence of the cross section.}
	\label{fig:ampSRW1}
\end{figure} 

The angle-wavelength radiation distribution $\frac{d^2{\cal U}}{d\theta d\lambda}$  and spectra simulated for the PU (with parameters listed in Tab.~\ref{und_params} appear in Fig.~\ref{fig:ampSRW1}(a,b). The radiation spectrum has a nonlinear correlation between wavelength and emission angle so that a large-aperture optical system, which is needed to avoid signal loss, will have a large bandwidth that is not transform limited. For instance the on-axis spectrum $\frac{d^2{\cal U}}{d\theta d\lambda}\big|_{\theta=0}$ has a full-width half-maximum (fwhm) relative bandwidth $\delta \lambda /{\lambda_0}\simeq 11$\% (with a maximum on-axis spectrum value at $\lambda_0=2.2$~\textmu{m}) while the angle-integrated spectrum integrated  $\int_0^{\theta=0.8/\gamma}\frac{d^2{\cal U}}{d\theta d\lambda} d\theta$ supports a fwhm bandwidth in excess of $\delta \lambda /\hat{\lambda}\simeq 32$\% with $\hat{\lambda}\simeq 2.4$~\textmu{m}; see  Fig.~\ref{fig:ampSRW1}(b).\\

 The bandwidth affects the amplification process via two predominant effects: (i) second order dispersion from the host medium and (ii) the finite spectral width of the amplifier gain. Field propagation to the kicker in SRW is performed in the Fourier-domain which makes it straight forward to include these effects. To account dispersion, the Sellmeier's equation for pure ZnSe~\cite{znse} is used to modify the phase of each Fourier (frequency) component passing through the amplifier, and we assume the presence of the $Cr2+$ ions does alter the dispersion significantly. Likewise accounting the wavelength dependence of $\sigma_s$ allows the use of Eq.~\ref{gain} to compute the amplitude growth of each Fourier component. The cross section $\sigma_s(\lambda)$ was modeled as the superposition of two Gaussian distributions $\sigma_s(\lambda)= \sum_{i=1}^2 \sigma_i \exp[-(\lambda-\mu_i)^2/(2w_i^2)$ with $\sigma_{1,2}=1.3\times10^{-18},$~$0.5\times10^{-18}cm$~$^2$, $\mu_{1,2}=2.07,$~$ 2.45$~\textmu{m} and $w_{1,2}=0.1,$~$0.25$~\textmu{m}; see Fig.~\ref{fig:ampSRW1}(c).

\begin{figure}[!h]
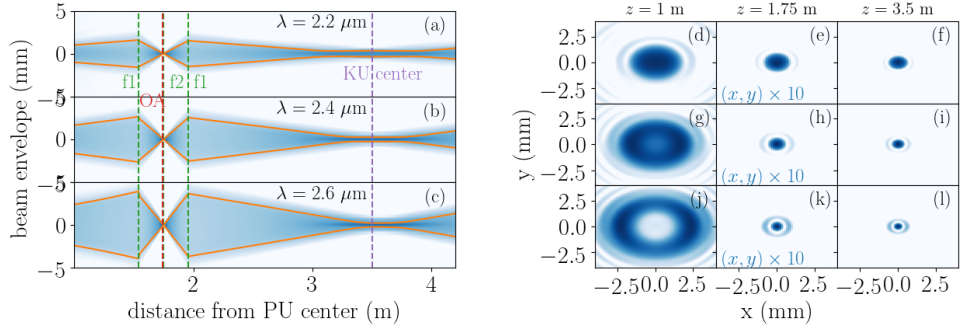

	\centering
	\includegraphics*[width=0.495\textwidth]{envelope.pdf}
	\includegraphics*[width=0.495\textwidth]{spots.pdf}
	\caption{Evolution of the beam envelope (orange trace) and transverse density (shaded blue) as function of propagation distance from PU center (a,b,c) for three radiation wavelength and corresponding radiation distribution for $\lambda=2.2$ (d-f), 2.4 (g-i), and 2.6~\textmu{m} (j-l). The density color map is logarithmic and normalized to the peak intensity (darker blue). The vertical lines indicate the location of the lenses (f1, f2), and optical amplifier (OA).The middle column has its coordinates scaled by a factor 10. }
	\label{fig:beamenv}
\end{figure}

The emitted undulator-radiation wave-packet is formed over the length of the PU and likewise, the energy exchange between the wave-packet and the particle occurs over the length of the KU. Ideally, the optical transport is set up so that the ABCD  matrix from the PU to KU centers is the identity matrix $\mathbf{I}$. Such a choice produces point-to-point imaging from a longitudinal location in the PU to its corresponding location in the KU. A set up providing the required point-to-point imaging consists of a three-lens telescope where two identical lenses with focal length $f_1$ are symmetrically located around a lens with focal length $f_2$; see Fig.~\ref{fig:oscinsert}. The focal lengths are given by  
\begin{equation}\label{eq:foclength}
f_1=\frac{d_1d_2}{d_1+d_2}\quad\quad\quad f_2=\frac{d_2^2}{2(d_1+d_2)},
\end{equation}
where $2(d_1+d_2)= 3.5$~m in IOTA is the distance from PU to KU centers; see Fig.~\ref{fig:oscinsert}. It should be noted that the point-to-point imaging allows for the telescope to be displaced longitudinally without introducing focusing errors. However, locating lens $f_2$ halfway between the PU and KU centers ensures the angular acceptance of the system is maximized for given lens diameters. The lens material was selected to be Barium Fluoride (BaF$_2$) owing to its small chromaticity~\cite{Refractive_index_HLI}. In addition to its optical imaging function, the optical system also focuses the PU radiation on the amplifier crystal (by choosing $f_1=184$~mm). SRW computations indicate the undulator-radiation spot radius to be $w_0 \simeq 100$~\textmu{m}. Figure~\ref{fig:beamenv}(a-c) shows the evolution of the beam envelope as a function of distance from the PU center for different wavelengths within a range where significant OA gain can be achieved. For these calculations, the lens parameters were selected to provide the required focal length according to Eq.~\ref{eq:foclength} for the nominal wavelength $\lambda=2.4$~\textmu{m}.  
One of the unconventional features associated with undulator radiation is the correlation between wavelength and divergence yielding a wavelength-dependent transverse distribution as showcased in  Fig.~\ref{fig:beamenv}(d-l). Given the needed pumping optical intensity of $I_p=125$~kW/cm$^2$, the required laser power is $P\simeq \pi w_0^2 I_p \sim 40$~W. Moreover, since $2(d_1+d_2)$ is fixed by real-estate constraints associated with the design of the IOTA accelerator, the selected focal length $f_1$ gives $f_2=13$~mm. 

Numerical simulations indicate that amplification of the PU-radiation pulse with a 7-dB gain produces a peak electric field $E_0\simeq 24.3$~V/m in the center of the KU (where the pulse is imaged). In comparison imaging of the original PU radiation pulse produces an electric field $E_0\simeq 10.9$~V/m in the absence of amplification; see Fig.~\ref{fig:ampSRW2}(a). In the later figure, we also compare the effect of amplification accounting for the limited bandwidth due to the optical-imaging system acceptance. For the 7-dB gain, we find that the pulse electric-field reduces to  $E_0\simeq 20.9$~V/m, when accounting for the decreased bandwidth, i.e. a $\sim 16$\% reduction compared to the ideal case. Additionally, the electric field is further reduced to  $E_0\simeq 19.1$~V/m when both the reduced bandwidth and host dispersion effects are taken into account; see Fig.~\ref{fig:ampSRW2}(a). \\
It should be finally pointed out that, owing to its large thermo-optics coefficient $\frac{dn}{dT}\simeq 70\times 10^{-6}$~K$^{-1}$, Cr:ZnSe is prone to thermal lensing. With an assumed pump intensity of 125 kW/cm$^2$ and taking the thermal conductivity $\kappa_t=1.0$~W/cm-K at 77~K~\cite{slack}, we find a thermal lensing focal length of $f_{th} \simeq  10$~mm~\cite{Kochner}. Therefore, the conceptual design depicted  in Fig.~\ref{fig:oscinsert} would have to be slightly altered by either introducing an additional compensating lens or curving the crystal surface to properly account for this effect and maintain the desired transfer matrix.
%
%
\section{Energy transfer from the amplified wave-packet to an electron}
Finally, we investigate the energy transfer from the amplified wave-packet to an electron in the KU (with magnetic field ${\pmb B}(z) =B_0\sin(\frac{2\pi}{\lambda_u} z)\hat{y}$ identical to the one generated by the PU ). We consider a test electron co-propagating with the previously computed wave-packet. The electron trajectory is taken to be in the KU mid-plane so that it can be described in the ($x,z)$ plane. The wave-packet field is parameterized as $E_x[x(t),z(t),t_o]$ where $x(t),z(t)$ are the particles coordinates in the KU and $t_o$ describes the relative arrival time between the electron and wave-packet in the KU.  Within the KU, the electron oscillates horizontally with velocity components~\cite{Wiedemann}
\begin{equation}
v_x(t)=-\beta c\frac{K}{\gamma}\sin(\omega_ut) \quad \quad 
v_z(t)=\beta c\bigg(1-\frac{K^2}{2\gamma^2}\sin^2(\omega_ut)\bigg)
\end{equation}
where $K\equiv \frac{eB_0}{mc^2k_u}$ is the undulator parameters,  $\omega_u\equiv ck_u$, and $\beta\equiv v/c=[1-\gamma^{-2}]^{1/2}$. The average longitudinal velocity associated with the electron while propagating in the KU is $\langle v_z\rangle=c\beta\left(1-K^2/4\gamma^2\right)$. Consequently, as the electron co-propagates with the wave-packet in the KU it will slip behind one optical cycle for every period of the KU. It is this condition that allows for a resonant interaction between the electron and wave-packet.

The rate of energy transferred to the particle is given by $\frac{d\mathcal{E}}{dt}=e\bm{v}\cdot\bm{E}$, resulting in a total energy exchange in the KU
\begin{equation}
\Delta \mathcal{E}=e\int E_x[x(t),z(t)]v_xdt.
\label{kick2}
\end{equation}
The latter integral can be evaluated numerically to yield the net energy gain experienced by the electron (here we take the initial delay between the electron and field so that the electron gains the maximum amount of energy). Theoretical expressions for the OSC energy kick in the absence of an optical amplifier can be found in Ref.~\cite{Andorf_PRAB}. It was especially shown that the energy kick grows considerably with the first lens aperture (which determines the bandwidth of the radiation accepted in the optical system). Our selected angular acceptance ($\gamma\theta_{max}=0.8$, corresponding to a lens radius of 10~mm) yields a kick amplitude of approximately 80\% of its theoretical maximum for the non-amplified case. 

\begin{figure}[!h]
	\centering
	\includegraphics*[width=0.495\textwidth]{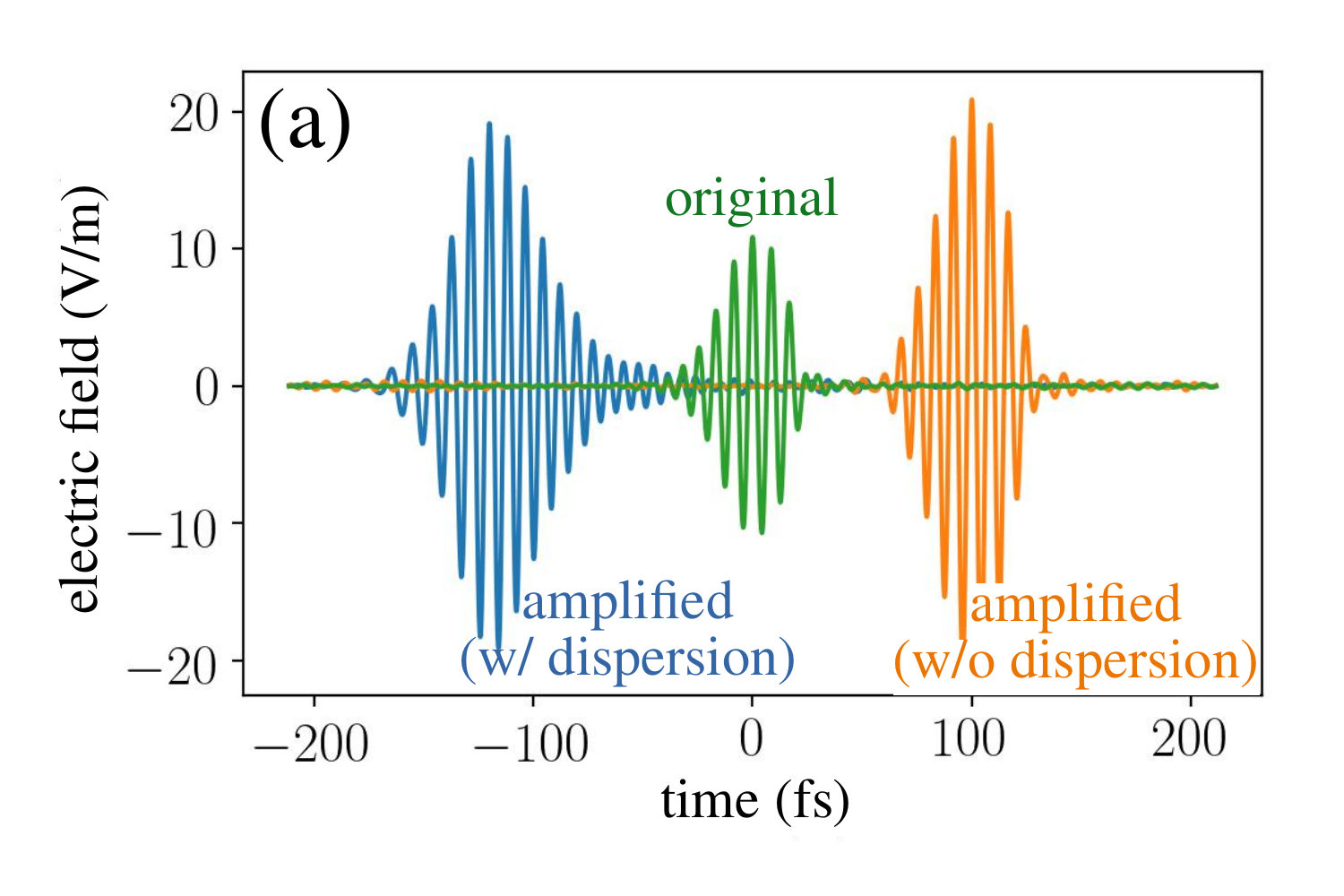}
	\includegraphics*[width=0.495\textwidth]{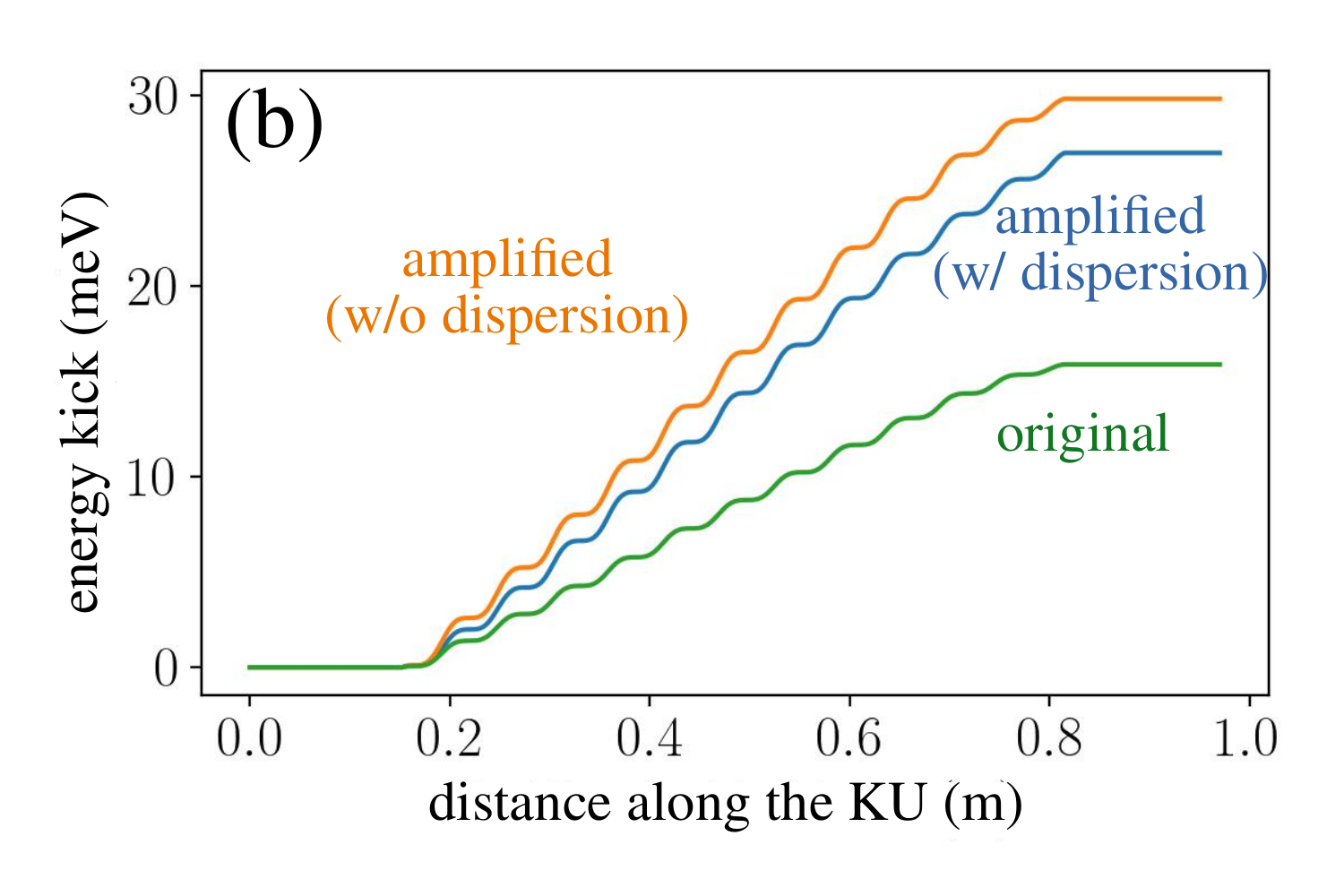}
	\caption{Electric field associated with the PU-radiation pulse imaged at the center of the KU (a) and corresponding energy gain experienced by an electron as it interacts with the pulse within the KU (b). The green, blue, and orange traces respectively correspond to the cases of the original PU pulse, and after amplification with and without accounting for the host dispersion. The finite bandwidth limited by the optical system aperture is taken into account in these calculations.  }
	\label{fig:ampSRW2}
\end{figure} 

Figure~\ref{fig:ampSRW2}(b) summarizes the energy exchange between an electron co-propagating with the numerically simulated PU radiation within the KU numerically. The calculations are performed for the proof-of-principle OSC experiment planned at IOTA considering a 1-mm-thick crystal. The electron initial phase is selected to yield the maximum final energy gain. For an ideal 7 dB amplifier (i.e. no dispersion and unlimited bandwidth) and a +$\mathbf{I}$ optical system with an angular acceptance of $\gamma\theta_{max}=0.8$, a  kick amplitude of 38.9~meV is computed.  However, the inclusion of both the host dispersion and spectral bandwidth reduces the energy-kick amplitude to 29.7~meV. This value is still a factor of 1.7 larger compared to the unamplified KU radiation which gives a kick of 17.4~meV and is sufficient to demonstrate the concept of active OSC in IOTA.

Our analysis is so far based on the room-temperature Cr:ZnSe absorption and emission cross-sections which have been extensively studied~\cite{Sorokina2,mirov}.  A recent investigation of the temperature dependence indicates that Cr:ZnSe’s performances moderately improve at liquid nitrogen temperatures~\cite{Riha}. However, the room-temperature spectra presented in Ref.~\cite{Riha} do not match other sources \cite{Sorokina2,mirov} making it problematic to assign cross-section values from their reported spectra. Nevertheless, based on the reported spectra we estimate that the gain associated with the amplifier discussed above would increase to $\sim 15$~dB yielding an overall kick amplitude of $\sim 40$~meV. However, given the discrepancy in the data from Ref.~\cite{Riha} with earlier work, we assume a conservative 7-dB gain value for the remainder of the paper. 
\section{Conclusion}
We have presented formulae for the single-pass gain of a Cr:ZnSe amplifier and used a wave-optics simulation model to compute the expected amplification of the broadband wave-packet emitted by a single electron passing through the PU. Furthermore, we find the amplifier gain is limited due to a saturable absorption effect of the pump laser, a problem exacerbated by the requirement of a small optical delay imposed on the amplifier to be compatible with the particle beam bypass chicane.
Solid-state gain mediums are an attractive approach for an OSC amplifier since, due to the small-signal input of the PU, the design is essentially independent of the accelerators' pulse structure (i.e. revolution period and charged-particle bunch duration). However, since the population inversion required for amplification is the result of the medium absorbing the pump laser, thermal effects and limits of the obtainable gain create a bottleneck. 

An alternative scheme based on an optical parametric amplifier (OPA) has also been considered for the amplification of undulator radiation~\cite{Pavlishin}. A significant obstacle to its deployment for OSC comes from the requirement that the OPA pump laser's pulse structure must match that of the accelerators.  In the case of a hadron collider, the bunches are typically on the order of a few nanoseconds in duration and have repetition rates exceeding MHz. Such a pulse format combined with the typical pump pulse intensity required to drive nonlinear mixing in OPAs [${\cal O}(\mbox{GW/cm$^2$})$] calls for mJ scale pulse energies corresponding to average pumping power on the order of kW's which is currently challenging to accomplish.

\section{Acknowledgments}
This work is supported by the U.S. Department of Energy under award No.DE-SC0018656 with Northern Illinois University and by the U.S. National Science Foundation under award PHY-1549132, the Center for Bright Beams. Fermilab is managed by the Fermi Research Alliance, LLC for the U.S. Department of Energy  Office of  Science  Contract  No.  DE-AC02-07CH11359.
\section{Disclosures}
The authors declare no conflicts of interest.


\end{document}